\documentstyle[aps,psfig,multicol]{revtex}
\input{epsf}
\newcommand{\bcols}{\ifpreprintsty\else\begin{multicols}{2}\fi}
\newcommand{\ecols}{\ifpreprintsty\else\end{multicols}\fi}
\begin{document}
\draft
\title{Decay of the Loschmidt Echo for quantum states with sub-Planck
scale structures}
\author{Ph. Jacquod, I. Adagideli, and C.W.J. Beenakker}
\address{Instituut-Lorentz, Universiteit Leiden, P.O. Box 9506, 2300 RA
Leiden, The Netherlands}
\date{\today}
\maketitle
\begin{abstract}
Quantum states extended over a large volume 
in phase space have oscillations from quantum interferences
in their Wigner distribution on scales smaller than $\hbar$
[W.H. Zurek, Nature {\bf 412}, 712 (2001)].
We investigate the influence of those sub-Planck scale structures
on the sensitivity to
an external perturbation of the state's time evolution.
While we do find an accelerated decay of the Loschmidt Echo
for an extended state in comparison to a localized wavepacket,
the acceleration is described entirely by the classical
Lyapunov exponent and hence cannot originate from quantum interference.
\end{abstract}
%\narrowtext
\bcols
One common interpretation of the Heisenberg uncertainty principle
is that phase-space structures on scales smaller than $\hbar$ 
have no observable consequence. The recent assertion of
Zurek \cite{Zurekn01} 
that sub-Planck scale structures in the Wigner function enhance the 
sensitivity of a quantum state to an external perturbation,
therefore, came out as particularly intriguing \cite{Albrecht01}
and even controversial \cite{Srednicki02}.
His argument can be summarized as follows. The 
overlap (squared amplitude of the scalar product) of
two quantum states $\psi$ and $\psi'$
is given by the phase-space integral of the product of their Wigner functions,
\begin{equation}
I_{\psi,\psi'} \equiv  |\langle \psi | \psi' \rangle |^2 = (2 \pi 
\hbar)^d \int  d{\bf r} d{\bf p} W_\psi W_{\psi'}.
\end{equation}
For an extended quantum state covering a large volume  
$A \gg \hbar^d$ of $2d$-dimensional phase space, 
the Wigner function $W_\psi$ exhibits oscillations from quantum interferences
on a scale 
corresponding to an action $\delta S \simeq \hbar^2/A^{1/d} \ll \hbar$. 
These sub-Planck scale oscillations are brought out of phase by a
shift $\delta p$, $\delta x$ with $\delta p \; \delta x \simeq \delta S 
\ll \hbar$. The shifted state $\psi'$ is then nearly orthogonal
to $\psi$ since $I_{\psi,\psi'} \approx 0$. Zurek concludes that
sub-Planck structures substantially enhance the sensitivity of a quantum state
to an external perturbation. 

A measure of this sensitivity is provided by 
the {\it Loschmidt Echo} \cite{Per84,Jal01}
\begin{equation}\label{fidelity}
M(t)=|\langle\psi|\exp(iHt)\exp(-iH_{0}t)|\psi\rangle|^{2},
\end{equation}
giving the decaying overlap of two wave functions that
start out identically and evolve under the action of two slightly different
Hamiltonians $H_0$ and $H=H_0+H_1$. (We set $\hbar=1$.) 
One can interpret $M(t)$ as the fidelity
with which a quantum state can be reconstructed by inverting the
dynamics with a perturbed Hamiltonian.
In the context of environment-induced dephasing,
$M(t)$ measures the decay of quantum interferences in
a system with few degrees of freedom interacting with
an environment (with many more degrees of 
freedom) \cite{Joos85}. 
In this case 
$\psi $ represents the state of the {\it environment}, which in
general extends over a large volume of phase space.
This motivated Karkuszewski, Jarzynski, and Zurek \cite{Zurek01}
to investigate the dependence of $M(t)$ on short-scale structures. 

In this paper we study the same problem as in Ref. \cite{Zurek01},
but arrive at opposite conclusions. Finer and finer structures
naturally develop in phase space when an initially narrow wavepacket
$\psi_0$ evolves in time under the influence of a chaotic Hamiltonian
$H_0$ \cite{Zurek01,Berman78}. As in Ref. \cite{Zurek01}, we observe
numerically a more rapid decay of $M(t)$ for $\psi = \exp(-iH_0 T) \psi_0$
as the preparation time $T$ is made larger and larger, with a saturation
at the Ehrenfest time. However, we demonstrate that this enhanced
decay is described entirely by the classical Lyapunov exponent, and hence
is insensitive to the quantum interference that leads to the sub-Planck 
scale structures in the Wigner function.

In the case of a narrow initial wavepacket,
$M(t)$ has been calculated semiclassically by Jalabert and 
Pastawski \cite{Jal01}. Before discussing extended states with 
short-scale structures, we recapitulate their calculation.
The time-evolution of a wavepacket centered
at ${\bf r}_0$ is approximated by
\begin{eqnarray}
\psi({\bf r},t) & = & \int d{\bf r}_0
\sum_s K_s^H({\bf r},{\bf r}_0;t) \psi_0({\bf r}_0), \\
K_s^H({\bf r},{\bf r}_0;t) & = & C_s^{1/2} 
\exp[i S_s^H({\bf r},{\bf r}_0;t)-i \pi \mu_s/2].
\end{eqnarray}
The semiclassical propagator is 
a sum over classical trajectories (labelled $s$) that
connect ${\bf r}$ and ${\bf r}_0$ in the time $t$. 
For each $s$, the partial propagator is expressed in terms
of the action integral $S_s^H({\bf r},{\bf r}_0;t)$ along $s$,
a Maslov index $\mu_s$ (which will drop out), and
the determinant $C_s$ of the monodromy matrix. 
After a stationary phase approximation, one gets
\begin{eqnarray}\label{semicl}
M(t) & \simeq & \bigl| \int d{\bf r}
\sum_s K_s^H({\bf r},{\bf r}_0;t)^* K_s^{H_0}({\bf r},{\bf r}_0;t)  
\bigr|^2.
\end{eqnarray}
Squaring the amplitude in Eq. (\ref{semicl}) leads to a double sum over
classical paths $s$, $s'$ 
and a double integration over final coordinates ${\bf r}$,
${\bf r}'$. Accordingly, $M(t)$ splits into diagonal 
($s=s'$, ${\bf r}={\bf r}'$) and nondiagonal 
($s \ne s'$ or ${\bf r} \ne {\bf r}'$) contributions. Since quantum 
phases entirely drop out of the diagonal contribution, its decay is
solely determined by the classical quantity $C_s \propto \exp(-\lambda t)$.
Here $\lambda$ is the
Lyapunov exponent of the classical chaotic dynamics, 
which we assume is the same for $H$ and $H_0$.
The nondiagonal contribution also leads to an exponential decay, which
however originates from the phase difference accumulated 
when travelling along a classical path with two different 
Hamiltonians \cite{Jal01}. The slope $\Gamma$ of this decay is 
the golden rule spreading width of an eigenstate of $H_0$ over the eigenbasis
of $H$ \cite{Jac01,Lew01}. 
Since $M(t)$ is given by the sum of these two exponentials, the Lyapunov
decay will prevail for $\Gamma > \lambda$. 

The Lyapunov decay sensitively depends on the choice of
an initial narrow wavepacket $\psi_0$ \cite{caveat}. The 
faster decay of $M(t)$ resulting from the increased complexity of
the initial state can be quantitatively investigated by considering 
prepared states $\psi = \exp(-i H_{0} T) \psi_0$, i.e. narrow 
wavepackets that propagate during a time
$T$ with the Hamiltonian $H_0$ \cite{caveat4}, 
thereby developing 
finer and finer structures in phase space as
$T$ increases \cite{Zurek01,Berman78}.
The stationary phase approximation to the fidelity then reads
\begin{equation}\label{semiclt}
M_T(t) = \bigl|\int d{\bf r}
\sum_{s} K_{s}^{H_\tau}({\bf r},{\bf r}_0;t+T)^* 
K_{s}^{ H_0}({\bf r},{\bf r}_0;t+T) \bigr|^2
\end{equation}
\noindent with the time-dependent Hamiltonian $H_\tau=
H_0$ for $\tau < T$ and $H_\tau=H$ for $\tau>T$. 

We can apply the same analysis as in Ref. \cite{Jal01} to the 
time-dependent Hamiltonian. Only the time interval $(T,t+T)$ of
length $t$ leads to a phase difference between $K_{s}^{H_\tau}$ 
and $K_{s}^{ H_0}$, because $H_\tau=
H_0$ for $\tau < T$. Hence the nondiagonal contribution to 
$M_T(t)$, which is entirely due to this phase difference,
still decays $\propto \exp(-\Gamma t)$, independent of the 
preparation time $T$.
We will see below that this is in agreement with a fully quantum mechanical
approach according to which
the golden rule decay is independent of the complexity of the initial
state.

The preparation does however have an effect on the 
diagonal contribution $M_T^{({\rm d})}(t)$ to the fidelity. It
decays $\propto \exp[-\lambda (t+T)]$ instead of $\propto \exp(-\lambda t)$,
provided $t$, $T \gg \lambda^{-1}$. This is most easily
seen from the expression
%\ecols
\begin{eqnarray}\label{semicld}
M^{({\rm d})}_T(t)&=&\int d{\bf r} \sum_{s} 
|K_{s}^{H_\tau}({\bf r},{\bf r}_0;t+T)|^2 \nonumber \\
&& \;\;\;\;\;\;\;\;\;\; \mbox{} \times 
|K_{s}^{H_0}({\bf r},{\bf r}_0;t+T)|^2, 
\end{eqnarray}
%\bcols
\noindent by following a path 
from its endpoint ${\bf r}$ to an intermediate point
${\bf r}_i$ reached after a time $t$. The time-evolution
from ${\bf r}$ to ${\bf r}_i$ leads to an exponential decrease
$\propto \exp(-\lambda t)$ as in Ref. \cite{Jal01}. Due to the classical
chaoticity of $H_0$, the subsequent evolution from
${\bf r}_i$ to ${\bf r}_0$ in a time $T$ brings in
an additional prefactor $\exp(-\lambda T)$. 

The combination of 
diagonal and nondiagonal contributions results in the bi-exponential
decay (valid for $\Gamma t$, $\lambda t$, $\lambda T \gg 1$)
\begin{equation}\label{prepdecay}
M_T(t) = A(t) \exp(-\Gamma t) + B(t) \exp[-\lambda (t+ 
T)],
\end{equation}
\noindent with prefactors $A$ and $B$ that depend algebraically on time.
The Lyapunov decay prevails if $\Gamma > \lambda$ and $t>\lambda 
T/(\Gamma - \lambda)$, while the golden rule decay dominates if either
$\Gamma < \lambda$ or $t < \lambda T/(\Gamma -\lambda)$.
In both regimes the decay saturates when $M_T$ has reached its 
minimal value $1/I$, where $I$ is the total
accessible volume of phase space in units of $\hbar^d$. 
In the Lyapunov regime, this saturation
occurs at $t=t_E-T$, where $t_{E}=\lambda^{-1} \ln I$ is the
Ehrenfest time. When the preparation time $T \rightarrow t_E$,
we have a complete decay within a time $\lambda^{-1}$ of the fidelity
down to its minimal value.

We now present numerical checks of these analytical results for
the Hamiltonian
\begin{equation}
H_{0}=(\pi/2\tau_0)S_{y}+(K/2S)S_{z}^{2}
\sum_{n}\delta(t-n\tau_0). \label{H0def}
\end{equation}
This kicked top model \cite{Haa00}
describes a vector spin of magnitude $S$ undergoing a free precession
around the $y$-axis and being periodically
perturbed (period $\tau_0$) by sudden rotations
around the $z$-axis over an angle proportional to $S_{z}$. 
The time evolution after $n$ periods is given 
by the $n$-th power of the Floquet operator
\begin{equation}
F_{0}=\exp[-i(K/2S)S_{z}^{2}]\exp[-i(\pi/2)S_{y}]. \label{Fdef}
\end{equation}
Depending on the kicking strength $K$, the classical dynamics is regular,
partially chaotic, or fully chaotic. 
We perturb the reversed time evolution by a
periodic rotation of constant angle around the $x$-axis, slightly
delayed with respect to the kicks in $H_0$,
\begin{equation}
H_{1}=\phi S_{x}\sum_{n}\delta(t-n\tau_0-\epsilon). \label{H1def}
\end{equation}
The corresponding Floquet operator is $F=\exp(-i\phi S_{x})F_{0}$. We set
$\tau_0=1$ for ease of notation.
We took $S=500$
(both $H$ and $H_{0}$ conserve the spin magnitude, the corresponding
phase space being the sphere of radius $S$) and calculated the
averaged decay $\overline{M}_T$ of 
$M_T(t=n)=|\langle\psi|(F^{\dagger})^{n}F_{0}^{n}|\psi\rangle|^{2}$ 
taken over 100 initial states.
 
We choose $\psi_0$ as a Gaussian wavepacket (coherent state) centered
on a point $(\theta,\varphi)$ in spherical coordinates.
The state is then prepared as $\psi = \exp(-i H_0 T) \psi_0$. 
We can reach the Lyapunov regime by 
selecting initial wavepackets centered 
in the chaotic region of the mixed phase
space for the Hamiltonian (\ref{H0def}) with kicking 
strength $K=3.9$ \cite{Jac01}.
Fig. 1 gives a clear confirmation of the predicted decay 
$\propto \exp[-\lambda (t+T)]$ in the Lyapunov regime.
The additional decay 
induced by the preparation time $T$ can be quantified via 
the time $t_c$ it takes for $\overline{M}_T$ to reach a given
threshold $M_c$ \cite{Zurek01}. We expect
\begin{equation}\label{tc}
t_c = -\lambda^{-1} \ln M_c-T,
\end{equation}
provided $M_c > 1/I=1/2S$ and $T < -\lambda^{-1} \ln M_c$.
In the inset to Fig. 1 we confirm this formula
for $M_c=10^{-2}$. As expected, $t_c$ saturates at the first kick 
($t_c=1$) when $T \simeq -\lambda^{-1} \ln M_c < t_E$. Numerical 
results qualitatively 
similar to those shown in the inset to Fig. 1 \cite{caveat2}
were obtained in 
Ref. \cite{Zurek01}, and interpreted there as the accelerated decay resulting
from sub-Planck scale structures. The fact that our numerical data is
described so well by Eq. (\ref{tc}) points to a classical rather than a
quantum origin of the decay acceleration. Indeed, Eq. (\ref{tc}) contains 
only the classical Lyapunov exponent as a system dependent parameter,
so that it cannot be sensitive to any fine structure in phase
space resulting from quantum interference.

\begin{figure}
\epsfxsize=3.3in
\epsffile{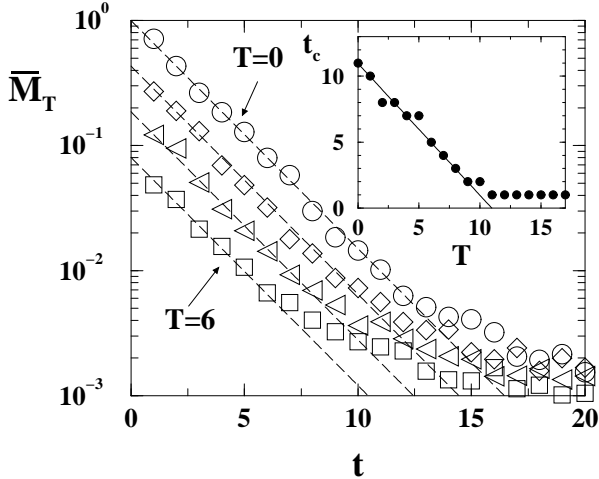}
\caption
{Decay of the average fidelity $\overline{M}_T$ for the kicked
top with parameters $\phi=1.2 \times 10^{-3}$,
$K=3.9$ and for preparation times
$T=0$ (circles), 2 (diamonds), 4 (triangles), and 6 (squares). In each case,
the dashed lines give the analytical decay 
$\overline{M}_T = \exp[-\lambda(t+T)]$,
in the Lyapunov regime with $\lambda=0.42$.
Inset: threshold time at which $\overline{M}_T(t_c) = M_c=10^{-2}$. 
The solid line gives the
analytical behavior $t_c = -\lambda^{-1} \ln M_c-T$.}
\label{fig:fig1}
\end{figure}

We next illustrate the independence of $M_T(t)$ 
on the preparation time $T$ in the golden rule regime, i.e. at larger
kicking strength $K$ when $\lambda > \Gamma$ \cite{Jac01}.
As shown in Fig. 2, the decay of $M_T(t)$ is the same 
for the four different preparation times $T=0$, 5, 10, and 20. 
We estimate the Ehrenfest time as $t_E \approx 7$,
so that increasing $T$ further does not increases the complexity of 
the initial state.

These numerical data give a clear confirmation of the 
semiclassical result (\ref{prepdecay}).
Previous investigations have established the existence of five 
different regimes for the decay of 
$M(t)$ \cite{Per84,Jal01,Jac01,Lew01,Cer01}, and since
only two of them are captured by the semiclassical approach used
above, we now show that short-scale structures do not affect
the remaining three. The five regimes correspond to different
decays: \\
(i) Parabolic decay,
$M(t) = 1-\sigma^2 t^2$, with
$\sigma^2 \equiv \langle\psi_0|H_1^2|\psi_0\rangle
-\langle\psi_0|H_1|\psi_0\rangle^2$, which
exists for any perturbation strength at short enough times. \\
(ii) Gaussian decay, $M(t) \propto 
\exp(-\sigma^2 t^2)$,
valid if $\sigma$ is much smaller than the 
level spacing $\Delta$. \\
(iii) Golden rule decay, $M(t) \propto \exp(-\Gamma t)$, with
$\Gamma \simeq \sigma^2/\Delta$, if $\Delta < \Gamma < \lambda$.\\
(iv) Lyapunov decay, $M(t) \propto \exp(-\lambda t)$, if $\lambda < \Gamma$.\\
(v) Gaussian decay, $M(t) \propto \exp(-B^2 t^2)$, if $H_1$
is so large thxdvat $\Gamma$ is  larger than the energy bandwidth $B$ of $H$. 

\begin{figure}
\epsfxsize=3.3in
\epsffile{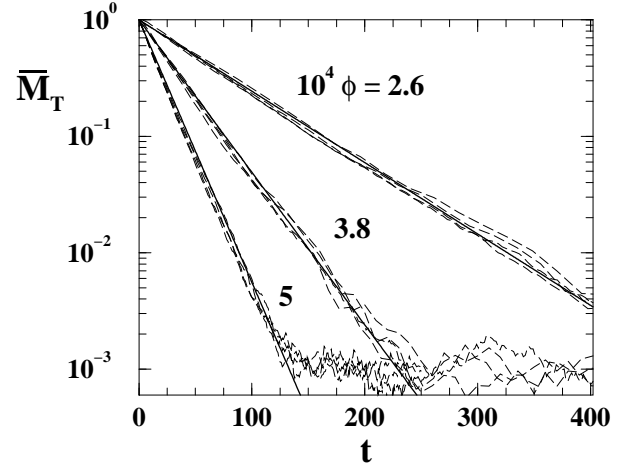}
\caption
{Decay of $\overline{M}_T$ in the golden rule regime for 
$\phi=2.6 \times 10^{-4}$,
$3.8 \times 10^{-4}$, $5 \times 10^{-4}$, $K=13.1$, and for 
preparation times $T=0$, 5, 10, and 20 (nearly indistinguishable dashed
lines). The solid lines give the corresponding 
golden rule decay with $\Gamma = 0.84 \; \phi^2 S^2$ \protect\cite{Jac01}.}
\label{fig:fig2}
\end{figure}

All these regimes except regime (iii) can be dealt with quantum mechanically
under the sole assumption that both $H_0$ and $H$ are classically
chaotic, using Random Matrix Theory (RMT) \cite{caveat3}.
Both sets of eigenstates 
$|\alpha\rangle$ of $H$ (with $N$ eigenvalues $\epsilon_\alpha$)
and $|\alpha_0 \rangle$ of $H_0$ (with $N$ eigenvalues $\epsilon_\alpha^0$)
are then rotationally invariant \cite{Mehta}. Expanding 
$\psi = \sum_\alpha \psi_\alpha | \alpha \rangle$ and assuming unbroken 
time-reversal symmetry, the fidelity (\ref{fidelity}) can be rewritten as
%\ecols
\begin{eqnarray}\label{expansion}
M(t) & = & \sum_{\alpha\beta\gamma\delta} \psi_\alpha \psi_\beta
 \psi_\gamma \psi_\delta
\langle \alpha |\exp(-i H_0 t) |\beta \rangle \nonumber \\
& & \;\;\;\; \times \; \langle \gamma |\exp(i H_0 t) |\delta \rangle 
\exp[i (\epsilon_{\alpha}-\epsilon_{\delta}) t].
\end{eqnarray}
%\bcols
RMT implies the $\psi$-independent average
$\overline{ \psi_\alpha \psi_\beta \psi_\gamma \psi_\delta} 
= (\delta_{\alpha\beta}\delta_{\gamma\delta}+
\delta_{\alpha\gamma}\delta_{\beta\delta}+\delta_{\alpha\delta} 
\delta_{\beta\gamma})/N^2$. The third contraction gives
a contribution $N^{-1}$ representing the saturation of
$M(t)$ for $t \rightarrow \infty$. The other two give the 
time dependence
\begin{equation}\label{tcontr}
\overline{M}(t) = N^{-1}+
2 N^{-2} \bigl|\sum_{\alpha\beta_0} |\langle \alpha | \beta_0 \rangle|^2
\exp[i (\epsilon_{\alpha}-\epsilon_{\beta}^0) t] \bigr|^2.
\end{equation}
For perturbatively weak $H_1$ one has
$\epsilon_{\alpha}=\epsilon_{\alpha}^0+\langle \alpha|H_1|\alpha\rangle$
and $\langle \alpha | \beta_0 \rangle = \delta_{\alpha,\beta_0}$. 
According to RMT
the matrix elements 
$\langle \alpha|H_1|\alpha\rangle$ are independent random numbers,
and for large $N$ the central limit theorem
leads to the Gaussian decay (ii) (or the parabolic decay (i) for short times).
At larger perturbation 
strength, $|\langle \alpha | \beta_0 \rangle|^2$ becomes Lorentzian,
\begin{equation}
|\langle \alpha | \beta_0 \rangle|^2=\frac{\Gamma/2 \pi}{ 
(\epsilon_{\alpha}-\epsilon_{\beta}^0)^2+\Gamma^2/4},
\end{equation}
with a width $\Gamma\simeq \overline{ |\langle \alpha_0 |H_1| 
\beta \rangle|^2}/\Delta$ given by the golden rule.
This leads to regime (iii).
Increasing $H_1$ further one obtains an ergodic distribution
$\overline{ |\langle \alpha | \beta_0 \rangle|^2}=N^{-1}$
and a straightforward calculation produces regime (v). This establishes that, 
under the sole assumption that $H_0$ and $H$ are 
classically chaotic, the decay of the fidelity in the 
three quantum regimes (ii), (iii), 
and (v) does not depend on the choice of the 
initial state $\psi$.

In summary, we have investigated the decay of the Loschmidt Echo,
Eq.\ (\ref{fidelity}), for quantum states $\psi = \exp(-i H_0 T) \psi_0$
that have spread over phase space for a time $T$. As in Ref. \cite{Zurek01},
we found a faster decay of $M_T(t)$ than for
a localized wavepacket, but only in the regime where the decay rate
is set by the classical Lyapunov exponent $\lambda$. Since quantum 
interferences play no role in this regime, we conclude that sub-Planck 
scale structures in the Wigner representation of $\psi$
do not influence the decay of the Loschmidt Echo.

This work was supported by the Swiss National Science Foundation and by the
Dutch Science Foundation NWO/FOM. We acknowledge helpful discussions with
N. Cerruti, R.A. Jalabert, and S. Tomsovic.

\ecols

\begin{references}

\bibitem{Zurekn01} W.H. Zurek, Nature {\bf 412}, 712 (2001).
\bibitem{Albrecht01} A. Albrecht, Nature {\bf 412}, 687 (2001).
\bibitem{Srednicki02} A. Jordan and M. Srednicki, quant-ph/0112139.
\bibitem{Per84} A. Peres, Phys.\ Rev.\ A {\bf 30}, 1610 (1984).
\bibitem{Jal01} R.A. Jalabert and H.M. Pastawski, Phys.\ Rev.\ Lett.\ {\bf
86}, 2490 (2001).
\bibitem{Joos85} E. Joos and H.D. Zeh, Z. Phys. B {\bf 59}, 223 (1985).
\bibitem{Zurek01} Z.P. Karkuszewski, C. Jarzynski, and W.H. Zurek,
quant-ph/0111002.
\bibitem{Berman78} G.P. Berman and G.M. Zaslavsky, Physica {A \bf 91}, 
450 (1978); M.V. Berry and N.L. Balasz, J. Phys. A {\bf 12}, 625 (1979).
\bibitem{Jac01} Ph. Jacquod, P.G. Silvestrov, and C.W.J. Beenakker,
Phys.\ Rev.\ E {\bf 64}, 055203 (R) (2001).
\bibitem{Lew01} F.M. Cucchietti, C.H. Lewenkopf, E.R. Mucciolo, H.M. 
Pastawski, and R.O. Vallejos, Phys. Rev. E {\bf 65}, 046209 (2002).
\bibitem{caveat} For example, if $\psi_0$ is
a coherent superposition of $N$ wavepackets, the diagonal
(Lyapunov) contribution is reduced by a factor $1/N$ while the
off-diagonal (golden rule) contribution remains the same.
\bibitem{caveat4} More generally, we could prepare the state
$\psi=\exp(-i H_p T) \psi_0$ with a chaotic Hamiltonian $H_p$
that is different from $H_0$ and $H$. We assume $H_p=H_0$
for ease of notation, but our results are straightforwardly extended to this
more general case.
\bibitem{Haa00} F. Haake, {\em Quantum Signatures of Chaos\/} (Springer,
Berlin, 2000).
\bibitem{caveat2} The similarity between the data in our Fig. 1
and in Ref. \cite{Zurek01} is only qualitative, mainly because of the much
larger value $M_c=0.9$ chosen in Ref. \cite{Zurek01}. For values of
$M_c$ close to 1, we expect that we can do perturbation theory
in $t$ which gives $M_T(t) = 1-\exp(\lambda T) \sigma^2 t^2$, 
and hence $t_c = \sqrt{1-M_c} \exp(-\lambda T/2)/\sigma$. Analyzing
the data presented in Fig. 2 of Ref. \cite{Zurek01} gives the values
$\sigma \approx 0.042$ and $\lambda \approx 0.247$. 
\bibitem{Cer01} N. Cerruti and S. Tomsovic, Phys.
Rev. Lett. {\bf 88}, 054103 (2002); T. Prosen and M. Znidaric, J. Phys. A
{\bf 35}, 1455 (2002); T. Prosen and T. Seligman, nlin.CD/0201038;
G. Benenti and G. Casati, Phys. Rev. E {\bf 65}, 066205 (2002);
D.A. Wisniacki and D. Cohen, nlin.CD/0111125.
\bibitem{caveat3} The RMT assumption relates 
the fidelity to the local spectral density of 
states. The Lyapunov regime is however not captured by this relationship,
see: D. Cohen, Phys. Rev. E {\bf 65}, 026218 (2001).
\bibitem{Mehta} M.L. Mehta, {\it Random Matrices} (Academic, New York, 1991).
\end{references}
\end{document}